\documentclass[useAMS,usenatbib]{mn2e}
\usepackage{times}
\usepackage{dcolumn}
\usepackage{graphicx}
\usepackage{aas_macros}

\newcommand{\msun}{\mbox{${\rm M}_{\odot}$ }}
\newcolumntype{d}{D{.}{.}{-1}}

\title[Optical spectroscopy of (candidate) ultra-compact X-ray
  binaries]{Optical spectroscopy of (candidate) ultra-compact X-ray
  binaries: constraints on the composition of the donor stars}

\author[Nelemans, Jonker \& Steeghs]{G.
  Nelemans$^{1,2}$\thanks{E-mail: nelemans@astro.ru.nl, based on
    observations made with ESO Telescopes at the Paranal Observatories
    under programmes 269.D-5026 and 073.D-0486} and
  P.G. Jonker$^{3,4,5}$ and D. Steeghs$^{4}$
  \\
  $^{1}$Department of Astrophysics, IMAPP, Radboud University
  Nijmegen, P.O. Box 9010, NL-6500 GL Nijmegen, The Netherlands\\
  $^{2}$Institute of Astronomy, University of Cambridge, Madingley Road, Cambridge CB3 0HA, UK\\
  $^{3}$SRON National Institute for Space Research, Sorbonnelaan 2,
  NL-3584 CA Utrecht, The Netherlands\\
  $^{4}$Harvard-Smithsonian Center for Astrophysics, 60 Garden Street,
  Cambridge MA 02138, USA\\
  $^{5}$Astronomical Institute, Utrecht University, P.O. Box 90000,
  3508 TA, Utrecht, The Netherlands\\
}

\begin{document}

\date{Accepted ... Received \today}

\pagerange{\pageref{firstpage}--\pageref{lastpage}} \pubyear{2006}

\maketitle

\label{firstpage}

\begin{abstract}
We present optical spectroscopy of several (candidate) ultra-compact
X-ray binaries (UCXBs) obtained with the ESO VLT and Gemini-North
telescopes. In only one of five observed UCXB candidates did we
find evidence for H in its spectrum (4U~1556-60). For XB~1905+00 the
optical counterpart is not detected. For the known UCXBs 4U~1626-67
and XB~1916-05 we find spectra consistent with a C/O and a He/N
accretion disc respectively, the latter is the first optical spectrum
of a He-rich donor in an UCXB. Interestingly, the C/O spectrum of
4U1626-67 shows both similarities as well as marked differences from
the optical C/O spectrum of 4U~0614+09. We obtained phase resolved
spectroscopy of 4U~0614+09 and the 44 min transient XTE~J0929-314. In
neither object were we able to detect clear orbital periodicities,
highlighting the difficulties of period determinations in UCXBs. We
reanalysed the spectra of XTE~J0929-314 that were taken close to the
peak of its 2003 X-ray outburst and do not confirm the detection of
H$\alpha$ emission as was claimed in the literature. The peak spectra
do show strong C or N emission around 4640\AA, as has also been
detected in other UCXBs.  We discuss the implications of our findings
for our understanding of the formation of UCXBs and the Galactic
population of UCXBs. At the moment all studied systems are consistent
with having white dwarf donors, the majority being C/O rich.
\end{abstract}


\section{Introduction}\label{introduction}

Low-mass X-ray binaries are systems in which a neutron star or black
hole accretes from a low-mass companion. Most systems have orbital
periods of hours to days and are consistent with the scenario
\citep{heu83} in which the donors are main sequence or evolved,
hydrogen-rich, stars. A few ultra-compact X-ray binaries (UCXBs) have
orbital periods below an hour and are so compact that the donor stars
cannot be main sequence stars, but instead must be hydrogen poor
\citep[e.g.][]{nrj86,vh95}. These systems have recently attracted 
attention for a number of reasons, such as the discovery of transient
UCXBs which harbour millisecond X-ray pulsars
\citep[e.g.][]{rss02,ms02a} and the fact that such systems are strong
sources of low-frequency gravitational wave radiation and will be
guaranteed sources for the joint ESA/NASA mission
LISA\footnote{\texttt{http://lisa.esa.int\newline \indent
http://lisa.jpl.nasa.gov}} \citep[e.g.][]{phi02,nel03b,nel05}.

One of the distinguishing properties of UCXBs is their optical
faintness in outburst, as expected for the small accretion discs in
these systems \citep{vm94}. Therefore it is only with the advent of 8m
class telescopes that high-quality optical spectra have become
available, and consequently much of our current knowledge is based on
X-ray observations and photometry at different wavelengths. Due to the
nature of the donor stars in UCXBs the disc material is expected to be
hydrogen poor providing an opportunity to distinguish UCXBs from other
low-mass X-ray binaries through optical or UV spectroscopy. There are
12 systems with known or suggested orbital periods, including systems
in globular clusters. Six more systems are identified based on
similarities with known systems in either their X-ray spectra
\citep{jpc00} or as a result of their optical faintness
\citep{bji+06}. For a short review of UCXBs see \citet{nj06}.

We have started a systematic program to obtain optical spectra of
(candidate) UCXBs using the ESO Very Large Telescope and the
Gemini-North telescope in order to test the proposed formation
channels for UCXBs and probe the interior structure of the donor stars
through determination of the chemical composition of the accretion
discs in these systems. The result of the first part of this projects
was the discovery of carbon-oxygen accretion discs in 4U~0614+09,
4U~1543-624 and possibly 2S~0918-549 \citep{njm+04}. In this paper we
report on the second set of observations, as well as on a collection
of other spectra obtained from the ESO archive. In
Sect.~\ref{observations} we briefly describe the observations and the
data reduction techniques. In Sect.~\ref{results} we will present the
results of our observations for the individual systems. We end the
paper with a discussion of the implications of our observations for
our understanding of UCXBs (Sect.~\ref{discussion}) and a summary of
our conclusions (Sect.~\ref{conclusions}).

\section{Observations and reduction}\label{observations}


\noindent \textbf{Spectroscopic observations of five (candidate) UCXBs
  }\\ Spectra were taken with the FORS2 spectrograph
  on UT4 of the 8m Very Large Telescope on Paranal in Chile. For each
  object we took spectra both with the 600B and 600RI holographic
  grisms, with a 1" slit, using 2x2 on-chip binning.  This setup
  resulted in coverage of 3326 -- 6359 \AA\ with mean dispersion of
  1.48 \AA/pix (resolution about 7\AA) for the 600B spectra and 5290
  -- 8620 \AA\ with mean dispersion of 1.63 \AA/pix (resolution about
  7\AA) for the 600RI spectra. Exposure times were 2754 s.  In
  addition VLT spectra of 4U1626-67, 4U~0614+09 and 4U~1556-60 were
  extracted from the ESO archive and reduced in the same way as the
  other spectra. 

\noindent \textbf{Spectroscopic DDT observations of XTE J0929-314}\\ Spectra were taken first with the UVES spectrograph
on the VLT . Unfortunately, the source had faded too much
(cf. Fig.~\ref{fig:0929_lc}). The spectra of the individual exposures
of 150 s could not be extracted so all 30 spectra were first combined
before the spectrum was extracted. Additional spectra were taken with
the FORS2 spectrograph with the 1200R and 1400V holographic grisms,
with a 1" slit, using 2x2 on-chip binning (resolutions about 3\AA) and
exposure times of 150 s. This setup resulted in coverage of 4580 --
6154 \AA\ with mean dispersion of 0.77 \AA/pix for the 1400V spectra
and 5869 -- 7369 \AA\ with mean dispersion of 0.73 \AA/pix for the
1200R spectra.  We also retrieved from the ESO archive the 3 spectra
that were taken close to the peak of the X-ray outburst with the ESO
3.6m telescope using the EFOSC spectrograph \citep{ccg+02}.

\noindent \textbf{Spectroscopic observations of 4U~0614+09 with Gemini}\\ Time resolved spectra were taken with the
 GMOS-North spectrograph on Gemini-North on Mauna Kea using the B1200
 grating with a 0.75'' slit (resolution about 2\AA) with 4x4 on-chip
 binning and an integration time of 293 s. A total of 52 spectra
 were taken between 30/10/2003 and 22/11/2003.

 For all of the above setups, data reduction was performed using
standard IRAF\footnote{IRAF is distributed by the National Optical
Astronomy Observatories} tasks.  The bias was removed using the
overscan region of the CCD, after which the images were flatfield
corrected using the standard calibration plan flatfields. Spectra were
extracted using optimal extraction \citep{hor86} with the
\texttt{apall} task. Arc lamp spectra were extracted from the same
place on the CCD. The 600B wavelength calibration was obtained using
the positions of 14-17 lines, giving a root-mean-square scatter of
less than 0.1 \AA\ in fitting a cubic spline.  The 600RI wavelength
calibration uses typically 40 lines and gives a root-mean-square
scatter of 0.15 \AA. The GMOS-North spectra were calibrated with 40
lines, giving an scatter of less than 0.08 \AA.

Because observations of spectroscopic standard stars were not obtained
close in time to our observations, the spectra were not flux
calibrated. The spectra of each object were averaged and the few
remaining cosmics removed by hand. The reduced spectra were
subsequently imported in the MOLLY package for further analysis. We
normalised the spectra by fitting a second order cubic spline to the
continuum at line-free regions of the spectrum.

\section{Results}\label{results}

In the following sections we will discuss the results for the
different objects in detail. Because there are very few optical
spectra of UCXBs known, interpretation of these spectra generally is
difficult. We found that simple LTE models are a good tool to identify
lines from the different possible elements that the discs of UCXBs can
contain, even though the LTE assumption is probably not justified
\citep{njm+04}. The LTE model that we use consists of a uniform slab
of gas, with constant density and temperature. For more details of
this model, see \citet{mhr91}and \citet{njm+04}. We discuss the interpretation of
our results and the applicability of our LTE modeling in
Sect.~\ref{discussion}.

\subsection{The known UCXB XB~1916-05}\label{1916}

\begin{table}
\caption{Most prominent lines in the model spectrum for 4U~1916-05}
\label{tab:1916}
\begin{center}
\begin{tabular}{ll} \hline
lines (\AA) & element \\ \hline
4097.36/ 4103.39 & N\textsc{iii} \\
4195.74/ 4200.07 & N\textsc{iii} \\
4471.68 & He\textsc{i} \\
4510.88/ 4510.96/ 4514.85/ 4518.14/ 4523.56 & N\textsc{iii}\\
4634.13/ 4640.64/ 4641.85/ 4641.96 & N\textsc{iii}\\
4686.75 & He\textsc{ii}\\
4858.70/ 4858.98/ 4861.27/ 4867.12/ 4867.17 & N\textsc{iii}\\
5015.68 & He\textsc{i}\\
5320.87/ 5327.19 & N\textsc{iii}\\
5666.63/ 5679.56 & N\textsc{ii}\\
5875.62 & He\,\textsc{i}\\
6559.71 & He\,\textsc{ii}\\
6610.56 & N\,\textsc{ii}\\
6678.15 & He\,\textsc{i}\\
7065.19 & He\,\textsc{i}\\ \hline
\end{tabular}
\end{center}
\end{table}

\begin{figure}
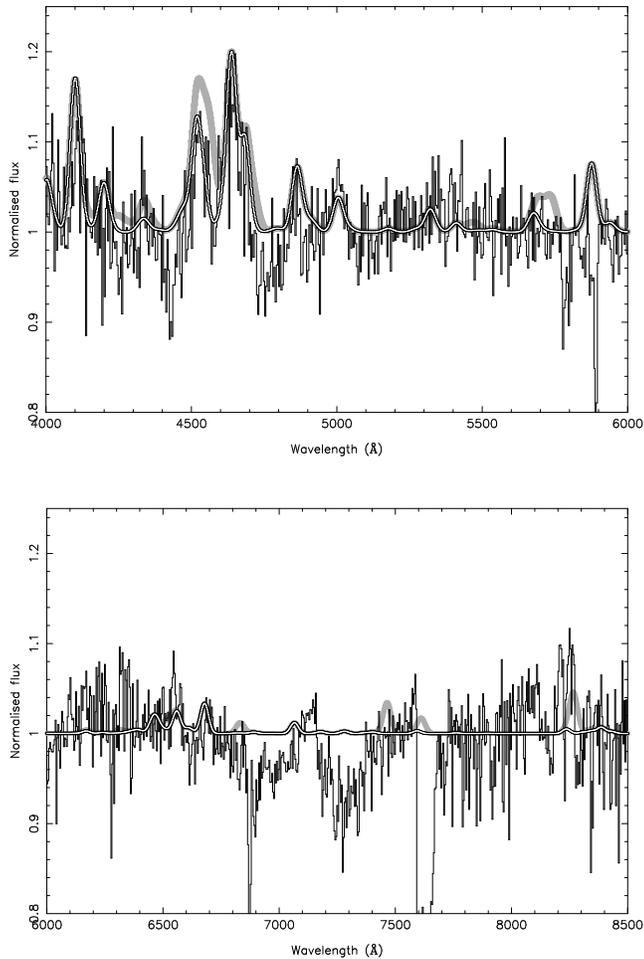

  \includegraphics[angle=-90,width=\columnwidth,clip]{1916_blue_new}

\vspace*{0.2cm}
  \includegraphics[angle=-90,width=\columnwidth,clip]{1916_red_new}
  \caption{Normalised (and rebinned) spectrum of the known UCXB
  XB~1916-05, showing broad features, with a 30 kK LTE model
  consisting of pure helium plus nitrogen (thin black/white line) and
  a model including heavier elements at solar composition (thick grey
  line). }
\label{fig:1916}
\end{figure}

With the VLT we obtained spectra of the known UCXB 4U 1916-05, which
has a 50 min orbital period \citep{wmc+82,ws82}. The faint optical
counterpart (V=21) has inhibited spectral studies until now. The
discovery of 270 Hz burst oscillations \citep{gcm+01} shows the
neutron star is spinning rapidly. Three spectra were taken with the
600RI grism, one with the 600B grism. In Fig.~\ref{fig:1916} we show
the normalised, average spectrum, which on first sight looks similar
to the spectrum of the C/O disc of 4U 0614+09 \citep{njm+04}, but in
addition broad emission around 4540\AA\ is present. We found a good
match with an LTE model consisting of pure helium plus overabundant
nitrogen (see Fig.~\ref{fig:1916}). The model has a temperature of
28,000 K, a particle density of $4 \times 10^{13}$~cm$^{-3}$ and a
line of sight through the medium of $10^7$~cm. The relative number of
He and N atoms is 90:3, which is about 10 times higher than material
of originally solar abundance which has been processed through the CNO
cycle. The most prominent lines are from He\textsc{i}, He\textsc{ii},
N\textsc{ii} and N\textsc{iii}, and are listed in
Table~\ref{tab:1916}. In the figure we also plot a model with the same
He and N abundance, but with the elements heavier than Ne at solar
abundance. Although there are some small differences, it is not
possible to determine the original metallicity of the donor. However,
if the high N abundance is true, an initially high metallicity is
likely.

We conclude that the detection of a He dominated accretion disk
spectrum provides the first direct evidence for a helium donor in an
UCXB. 

\subsection{The known UCXB 4U 1626-67 compared to 4U 0614-09}\label{1626_0614}

\begin{figure}
  \includegraphics[angle=-90,width=\columnwidth]{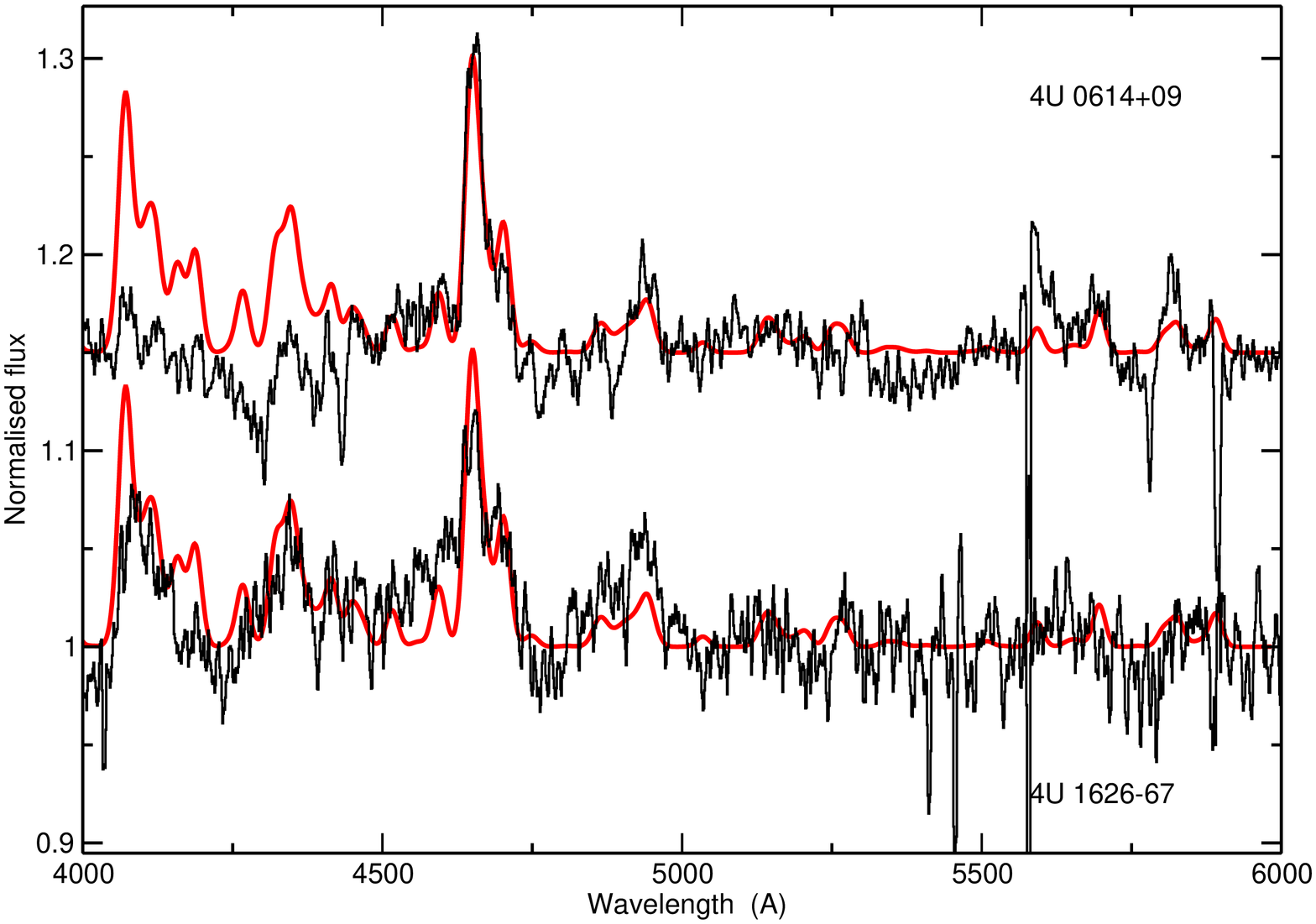}
  \includegraphics[angle=-90,width=\columnwidth]{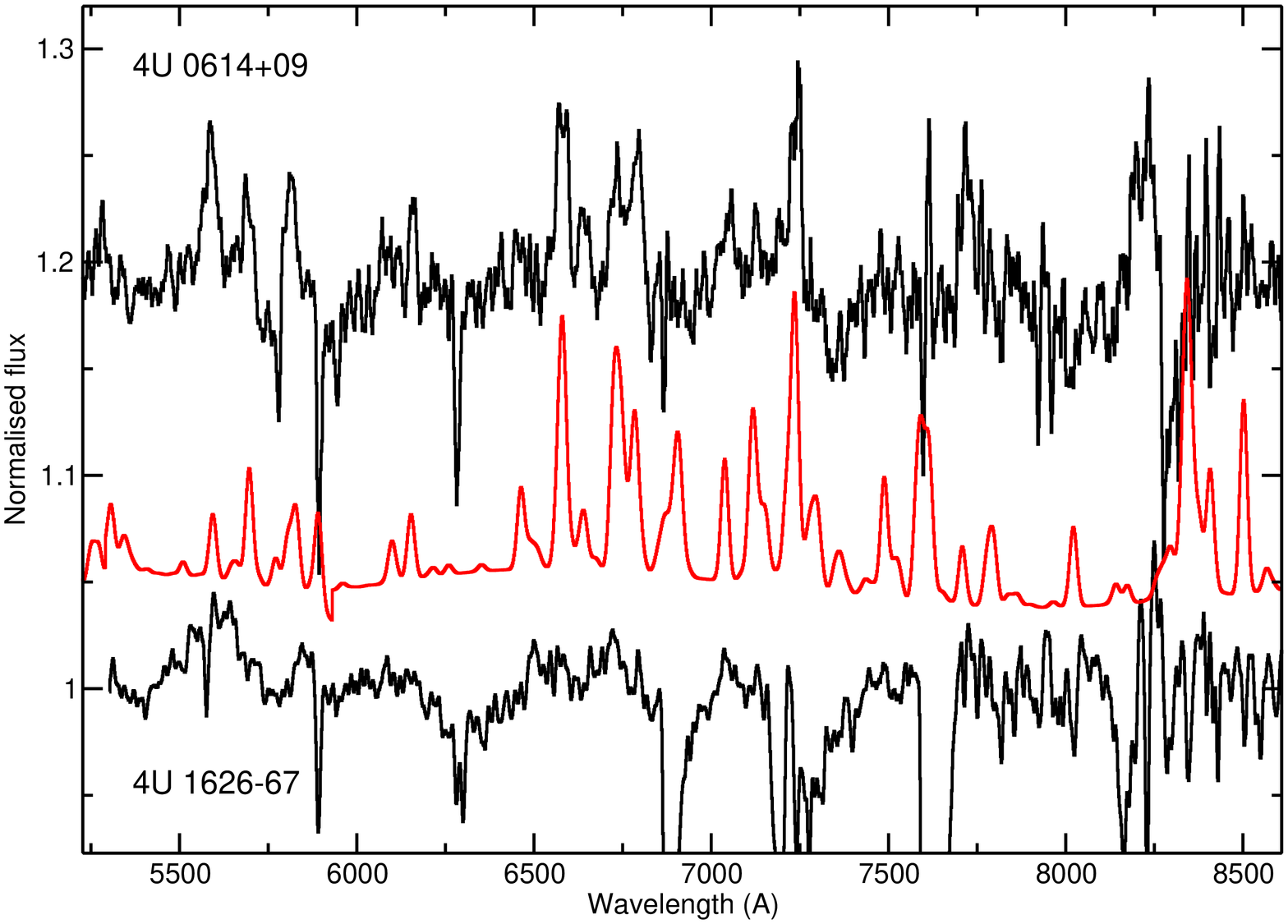}
  \caption{Normalised spectra of 4U~1626-67 and 4U~0614+09. Below
4500\AA\ the spectrum of 4U~1626-67 follows the LTE model of a C/O disc
that fitted our previous spectra of 4U~0614+09 (with wavelengths above
4600\AA) much better than 4U~0614+09 itself. In the red part of the
spectrum we plot the model which we used for the line identification
of 4U~0614+09 in the middle. Most features are C features and are
weaker in 4U~1626-67.}
\label{fig:1626_0614}
\end{figure}

We obtained spectra of the 42 min. binary 4U 1626-67, which harbours a
7 s X-ray pulsar \citep{mmn81}, making it a very different system to
the other UCXBs, which all have old, low-field neutron stars as
accretors. For a discussion of the formation of this system see
\citet{ynh02} and references therein. The system shows strong line
emission in the X-ray spectrum, which has been identified with O and
Ne lines \citep{sch+01}. Its UV spectrum shows strong carbon and
oxygen lines \citep{haw+02}. We therefore expect the optical spectrum
to show C/O lines, just as the C/O disc of 4U~0614+09 \citep{njm+04}.

\begin{figure}
  \includegraphics[angle=-90,width=\columnwidth,clip]{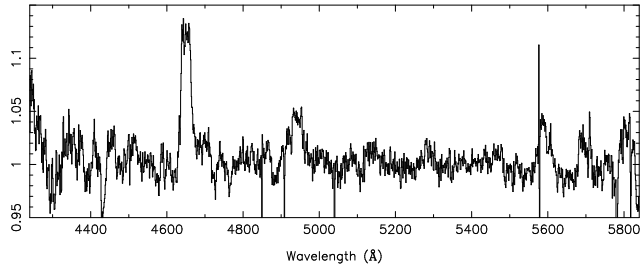}
 \caption[]{Average normalised spectrum of 4U~0614+09, taken with GMOS
 on Gemini-North. }
\label{fig:0614_GMOS_aver}
\end{figure}

Three spectra were taken with the 600RI grism, one with the 600B
grism.  In addition we retrieved two spectra taken with the 600B grism
plus four spectra taken with the 600R grism (exposure time 1735 s),
and three spectra taken with the 1400V grism (exposure time 1100 s)
from the ESO archive. In \citet{nj04} we already showed a comparison
between the spectrum of 4U 1626-67 and our earlier 4U~0614+09 spectrum
which looked very much like each other, in agreement with the
interpretation of the donors of both these systems being C/O
rich. Indeed 4U 1626-67 follows the simple C/O LTE model that we made
for 4U 0614+09 reasonably well even at bluer wavelengths than our 2003
4U~0614+09 spectrum reached. However, a second spectrum of 4U~0614+09,
obtained with the VLT with the same setup and wavelength coverage as
our 4U~1626-67 spectrum \citep[that we extracted from the ESO archive,
see also][]{wnr+06} clearly does \emph{not} follow the simple LTE
model!  (see Fig.~\ref{fig:1626_0614}). The blue part of the
4U~0614+09 spectrum does not show the prominent (O\textsc{II})
features \citep[see also][]{wnr+06}.  The small plateau just blue ward
of the 4640\AA\ feature in 4U~0614+09 seems a result of the continuum
normalisation (cf. Fig.~\ref{fig:0614_GMOS_aver}).

In Fig.~\ref{fig:0614_GMOS_aver} we show the average spectrum of
4U~0614+09 from all 52 spectra taken with GMOS on Gemini-North. The
spectrum shows the same features as the VLT spectra
(Fig.~\ref{fig:1626_0614}). We used the GMOS spectra to look for
variability in the line profiles on short timescales in order to
search for signals of the orbital period. We produced periodograms of
the data for the range of expected periods, but no periodic signals
were found.

\begin{figure}
  \includegraphics[width=\columnwidth,angle=-90,clip]{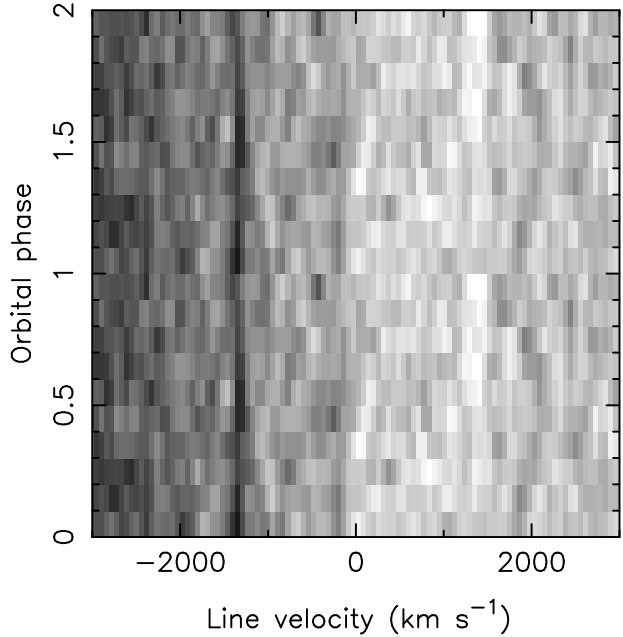}
 \caption[]{Trailed spectra of 4U~0614+09 around 4930\AA\ folded on a
 period of 48.547 min., showing a sinusoidal velocity variation in
 weak line around 4960\AA\ (CI 4959?) visible at $\sim$+2000
 km~s$^{-1}$. The amplitude of the variation is 190 km~s$^{-1}$, which
 implies an inclination of about 15$^\circ$ for a neutron star mass of
 1.4 \msun. The feature around $-1700$ km~s$^{-1}$ is the result of an
 artifact in the CCD. }
\label{fig:0614_folded}
\end{figure}

Another way to pick up coherent patterns in a variable environment is
to search by eye. We used a utility (kindly provided to us by Gijs
Roelofs) which shows movies of trailed spectrograms, folded at slowly
varying periods. In this way we browsed through the data, without
finding a clear periodic signature. The most convincing possible
period is 48.547 min, which produces a rather nice sinusoidal pattern
in a weak absorption line around 4960\AA\ (possibly CI 4958.7\AA), as
shown in Fig.~\ref{fig:0614_folded} (at velocity +2000
km~s$^{-1}$). However, the significance of this ``detection'' is
marginal at best and has been found after many hundreds of trial
periods. Interestingly, this period fits in rather well with the
period of 50 min. suggested recently based on photometric ULTRACAM
data \citep{ojd+05}. If these periods are true the latter period would
be a super-hump period, just above the orbital period. The amplitude
of the velocity variation is 190 km~s$^{-1}$, which if it originates
in the donor and with the assumption of a $\sim50$ min orbital period
gives a mass function of 0.025. Assuming a neutron star mass of 1.4
\msun and a companion star that has a much lower mass, this implies an
inclination of 15$^\circ$ which is rather low.

\subsection{XTE~J0929-314}\label{0929}

\begin{figure}
 \includegraphics[width=0.9\columnwidth]{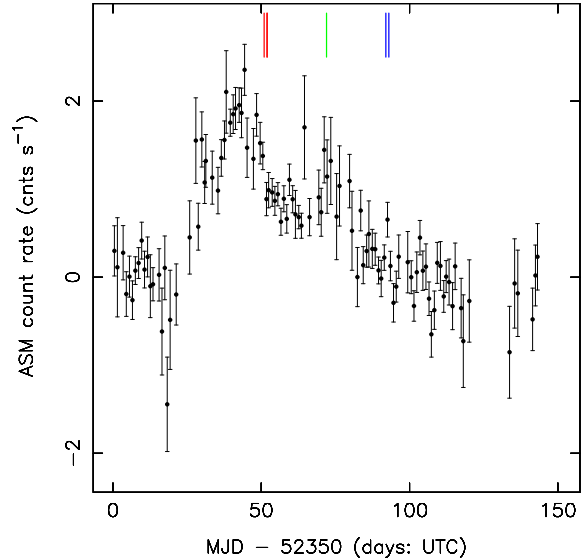}
\caption[]{X-ray light curve of XTE J0929-314 from the RXTE All sky
  monitor. The dates of the optical observations are indicated at the
  top of the plot (left to right): ESO 3.6m EFOSC observations, VLT
  UVES observations and VLT FORS2 observations.}
\label{fig:0929_lc}
\end{figure}

\begin{figure}
\begin{center}
  \includegraphics[angle=-90,width=\columnwidth,clip]{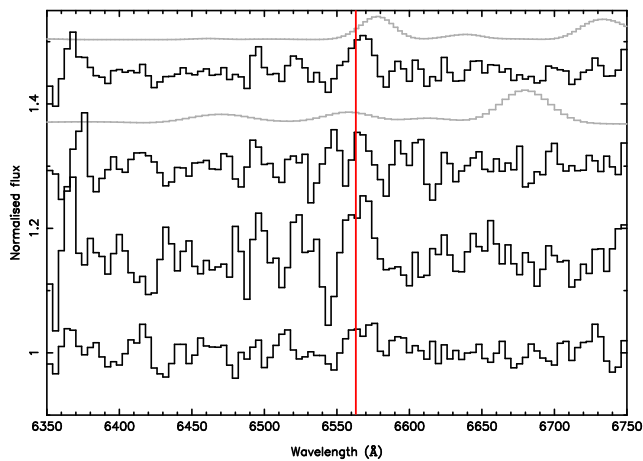}
  \caption{Region around H$\alpha$ of the outburst spectrum of XTE
  J0929-314 on the basis of which \citet{ccg+02} claim H$\alpha$
  emission which would contradict the interpretation of the donors as
  white dwarfs. The 3 individual spectra are shown at the bottom, the
  average as the top spectrum. The LTE models shown in
  Fig.~\ref{fig:0929} are shown in grey.}
\label{fig:0929_Ha}
\end{center}
\end{figure}

\begin{figure}
\begin{center}
  \includegraphics[angle=-90,width=\columnwidth]{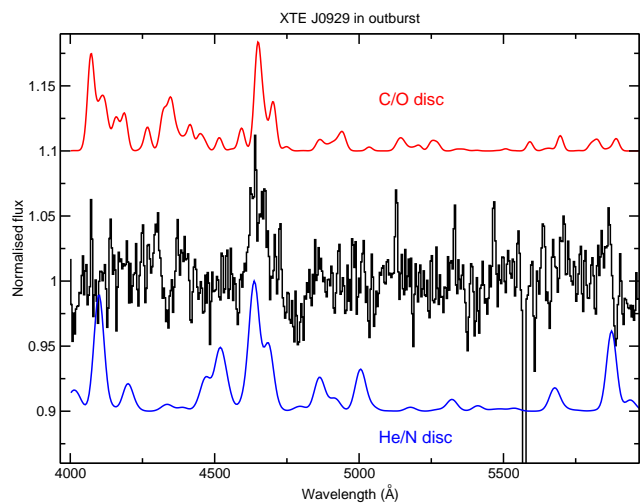}
  \caption{Spectrum of XTE J0929-314 taken close to the peak of the
  X-ray outburst. Also shown are LTE models for a C/O (top) and a He/N
  (bottom) disc are shown.}
\label{fig:0929}
\end{center}
\end{figure}

 XTE J0929-314 is one of the three transient UCXBs and was discovered
in 2002 \citep{rss02}. It has an orbital period of 44 min
\citep{gmr+02}. The X-ray lightcurve is shown in
Fig.~\ref{fig:0929_lc}. The system showed X-ray pulsations at 185 Hz
during the outburst \citep{rss02}. Near the peak of the X-ray outburst
a spectrum was taken with the ESO 3.6m telescope in which emission at
H$\alpha$ and around 4650\AA\ was reported \citep{ccg+02}. In
particular the H$\alpha$ emission is important, since this would
clearly point to a donor which still has some hydrogen left (see
Sect.~\ref{discussion}). However, the evidence for H$\alpha$ emission
is rather weak as is shown in Fig.~\ref{fig:0929_Ha}, where we display
the spectra. The formal significance of the H$\alpha$ emission is
3$\sigma$, but one has to keep in mind that there are many emission
lines in the night sky around H$\alpha$ that could lead to systematic
errors. There is clearly significant emission around 4650\AA\, as seen
in Fig.~\ref{fig:0929}. This is reminiscent of the strong emission
seen in the spectrum of the C/O accretion disc of 4U~0614+09
\citep{njm+04}, but also similar to the He + N spectrum of 4U~1916-05
discussed above. Because the peak spectra of XTE~J0929-314 are not of
very high quality it is difficult to distinguish, so as bracketing
models, we plot in Fig.~\ref{fig:0929} LTE models of a C/O disc (with
the same parameters as 4U~0614+09: a 27,000 K slab of 30 per cent C
and 70 per cent O) and a disc consisting of He and N with the same
parameters as 4U~1916-05 (3 per cent N by number).  The spectrum is
not good enough to determine the composition, although the lack of
emission around 4500\AA\ suggests a C/O donor. Between 4250 and
4400\AA\ there are hints of emission, and possibly absorption around
4300\AA. However, these features are not significant.

In order to see if a more quantitative criterion can be used we
determined the equivalent widths (EWs) of the strongest line complex
in our spectra: the range 4500-4700\AA, which includes the strong
N\textsc{III} feature around 4515\AA\ seen in 4U~1916-05. The EW of
4U~1916-05 is 15\AA, while the EWs of the other systems are:
4U~0614+09, 4\AA\ for the GMOS spectrum and 9\AA\ for the VLT
spectrum, 4U~1626-67 10 and 11 for the two VLT spectra and for
XTE~J0929-314 again 4\AA. Formal errors on these measurements are all
less than 0.5\AA. Although 4U~1916-05 has the largest EW in the wide
range, there is no clear gap between the EW values of C/O and He/N
discs. And we note that close inspection of the three spectra we have
of 4U~0614+09 show that differences in continuum normalisation cause
large systematic errors on the obtained EWs, thus hampering this
method. Despite all this, based on the EWs we again tentatively
classify XTE~J0929-314 as a C/O system.

After the discovery of XTE~J0929-314 we requested VLT DDT observations
to study the transient during the outburst (see Fig.~\ref{fig:0929_lc}
for the X-ray light curve with the times of our optical observations
indicated). We first attempted UVES observations (on 27 May 2002), but
it became soon clear that the target had faded too much. We therefore
changed to FORS2 observations. Unfortunately when the observations
were done (16/17 June 2002) the source again had faded and the spectra
show no clear features. From \citet{ggh+05} we derive the I-band
magnitudes on the dates of our observations as 18.5 (EFOSC), 19.5
(UVES), 20.1 (FORS2).

\subsection{4U 1556-60}\label{1556}

\begin{figure}
  \includegraphics[angle=-90,width=\columnwidth,clip]{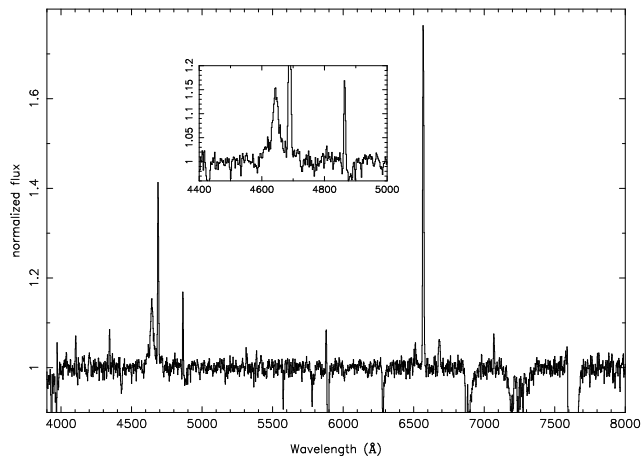}
  \caption{Normalised spectrum of 4U 1556-60, showing strong H and
  He\textsc{ii} emission, plus the Bowen blend at 4640\AA.}
\label{fig:1556}
\end{figure}

The source 4U~1556-60 was selected as a candidate UCXB based on the
suggested ``Ne feature'' in its X-ray spectrum \citep{ffm+03}. This
feature was discussed in the context of finding UCXBs by
\citet{jpc00}, who noted a similarity in the X-ray spectra of several
X-ray binaries, among which the known 20 min UCXB 4U~1850-087. They
interpreted this feature around 0.7 keV as an indication of enhanced
Ne in the transferred material.  In addition, earlier optical spectra
of 4U~1556-60 \citep{chc88,mpm+89} show strong emission between 4600
and 4700\AA, but no H$\beta$ and only weak emission around H$\alpha$,
similar to our spectrum of 4U~0614+09 (except of course for the
H$\alpha$).  We obtained four 600B and one 600RI spectra for this
object.  In Fig.~\ref{fig:1556} we show the normalised, averaged VLT
spectrum of 4U~1556-60 which shows a classical low-mass X-ray binary
spectrum with strong Balmer lines (4101, 4340, 4961 and 6563 \AA) and
lines from He\textsc{ii} (4686 (very strong), 5411 and 6678 \AA). The
equivalent width of the H$\alpha$ line is 7.8\AA\ and its FWHM is 540
km~s$^{-1}$. There is also strong emission at the Bowen blend, a C and
N complex around 4640\AA\ that is driven by He fluorescence. This
system thus probably is not an UCXB, suggesting that the ``Ne
feature'' is not a unique property of UCXBs.

\subsection{4U 1822-00}\label{1822}

The optical counterpart of 4U 1822-000 is a V=22 object \citep{ci85}.
  Due to its relatively high Galactic latitude (b=5.8 degrees) the
  extinction towards the source is low.  Even for a distance of 20
  kpc, the absolute magnitude would be 5.5. Such a faint absolute
  magnitude suggests a short orbital period, because the optical
  emission originates in an irradiated disc for which the brightness
  depends mainly on the \emph{size} of the disc, which scales with the
  orbital period \citep{vm94}. We obtained three 600B and four 600RI
  spectra, but due to the faintness of 4U~1822-00 its spectrum is of
  low quality. However, just as with 2S~0918-549 \citep{njm+04}, the
  spectrum does not show hydrogen or helium lines, making it clearly
  different from the spectra of hydrogen rich systems. Provisionally
  we classify this system as an UCXB, but with the current results not
  more can be said about the nature of the donor.

\subsection{XB 1905+00}\label{1905}

With an optical counterpart of V=20.5 and distance of $\sim$8kpc,
  i.e. M$_{\rm V}$= 4.9 \citep{ci85}, this source is again at the 
  faint end of the absolute magnitude distribution and thus a good
  candidate. The spectrum of XB 1905+00 shows the standard features of
  an early G star. As already discussed in \citet{nj04}, this puzzling
  result is possibly due to a chance alignment. The acquisition image
  of the object obtained with a seeing of 0.6 arcsec suggest the
  source actually is a blend of two objects. A detailed discussion of
  recent X-ray and optical observations suggests that the optical
  counterpart has faded beyond detection and neither of the objects is
  the actual counterpart \citep{jbn+06}.

\section{Discussion}\label{discussion}

\begin{table}
\caption[]{Overview of some properties of UCXB (candidates). Data from
  \citet{nj06} and references therein.}
\begin{center}
\begin{tabular}{llll}\hline
Name & Period & Persistent/ & Spec  \\
 & (min) & Transient&   \\
\hline
4U 1543-624   & 18(?)& P &  C/O \\ 
XTE J1807-294 & 40   & T & - \\
4U 1626-67    & 42   & P & C/O \\
XTE J1751-305 & 42   & T & - \\
XTE J0929-314 & 44   & T & C/O? \\
4U 1916-05    & 50   & P & He/N \\ 
4U 0614+09    & 50?  & P & C/O \\ 
2S 0918-549   & ?    & P & He? \\
4U 1822-00    & ?    & P & ? \\
XB 1905+000   & ?    & T & - \\ \hline
\end{tabular}
\end{center}
\label{tab:properties}
\end{table}

We have, by now, collected spectra of seven UCXB (candidates) in the
field, which we list in Table~\ref{tab:properties}. We also list the
chemical composition we have derived from these spectra. We obtained
the first optical spectrum of a He transferring UCXB,
XB~1916-05. However, the spectrum does not look at all like we had
anticipated, i.e. with strong helium emission lines, like the spectra
of AM CVn stars ES Cet and GP Com \citep[e.g.][]{ww02,mms+03}. This
means that really high S/N spectra are needed to determine the
chemical composition of the donors and that our conclusion in
\citet{njm+04} that 2S~0918-549 most likely has a C/O donor due to the
lack of strong helium emission lines was premature. Indeed,
\citet{icv05} suggest a helium donor is more likely because of the
properties of its X-ray bursts. It does show, however, that if there
is He in the disc, it shows up in the lines, so we are rather
confident that our earlier conclusions about the existence of pure C/O
discs still hold. Indeed from the LTE models, but also from recent
efforts modelling NLTE disc spectra \citep{wnd+04,wnr+06} it is
unlikely that H or He are hidden in these C/O discs.

In summary we have obtained spectra of 7 UCXB (candidates) and have
found 3 cases of a C/O donor (plus one, possibly), one of a He/N donor
(plus one, possibly), and two cases where we cannot determine the
chemical composition. Before turning to the interpretation of these
results we discuss some caveats in the determination of these
compositions.

\subsection{LTE modelling}\label{LTE}

Our chemical composition results are based on modelling of the
accretion disc as very simple, single temperature, homogeneous, slabs
of material for which we calculate the LTE spectrum. In real life
these accretion discs are likely multi-temperature and NLTE effects
almost certainly are important. Even worse, these discs are irradiated
by the strong X-ray emission originating around the neutron
star. Therefore it is surprising that the LTE models reproduce the
observed spectra reasonably well. Indeed, contrary to what one would
expect for these strongly irradiated discs, we see remarkably little
evidence for highly ionised species of the elements which would be
produced by photo-ionisation. Recently efforts have started to produce
more realistic, NLTE spectral models of UCXBs
\citep{wnd+04,wnr+06}. Unfortunately, the current models, although
reproducing the observed spectra qualitatively do not provide
quantitative results as it currently is not possible to fit all
observed features simultaneously, either due to incompleteness in the
radiation transfer physics or in the accretion disc physics. Until
these discrepancies are resolved, we have to live with significant
uncertainties in the derived chemical compositions.

\subsection{Formation of UCXBs and their Galactic population}\label{formation}

The main aim of this investigation is to use the determined chemical
composition of the donor stars in UCXBs to constrain the possible
formation channels and to come to a better understanding of the
evolution of ultra-compact binaries and their Galactic population.

For a detailed discussion of the formation of ultra-compact binaries
in the field we refer to \citet{nrj86,npv+00,prp02} and references
therein. In short there are three routes, differentiated by the nature
of the donor star: (i) a white dwarf donor when a detached binary with
a white dwarf and a compact object comes into contact due to angular
momentum losses via gravitational-wave radiation; (ii) a
(semi-degenerate) helium star donor that evolved from a helium core
burning star that filled its Roche lobe to a compact object and (iii)
the core of a star that filled its Roche lobe to a compact object at
the end of the main sequence and thus has a helium rich core
\citep[see also][]{svp05a}. In globular clusters the formation of
X-ray binaries is probably dominated by dynamical interactions
\citep[see][]{ver04}. Based on the inferred mass transfer rated of
4U~1626-67 and 4U~1916-05 \citet{nrj86} suggested evolved secondaries
as donors in these systems.

These different formation scenarios result in principle in different
chemical compositions which is one of the reasons for our study, but
there is some overlap. The three formation scenarios will yield the
following chemical composition:

\begin{itemize}
\item White dwarf donors: depending on the nature of the white dwarf
  the transferred material will be mainly He with CNO processed
  (i.e. mainly N) material for a He-core white dwarf, or a C/O mixture
  in case the donor is a C/O-core white dwarf.
\item Helium star donor: He with little N, plus possibly helium
  burning products (i.e. C and O) depending on the amount of helium
  burning that has taken place before the donors fills its Roche lobe.
\item Evolved secondaries: He plus CNO processed material and,
  depending on the exact evolutionary history and the phase of the
  evolution, some H.
\end{itemize}

In this light we can interpret the observed UCXBs.  The possible
presence of H in the peak spectrum of XTE~J0929-314 is very important
as only the third formation scenario can comfortable explain
this. White dwarfs and helium stars do have a thin H envelope before
they fill their Roche lobes, but this is typically expected to be at
most 0.01 \msun (likely less, but rather uncertain if the preceding
evolution was through a common-envelope phase) and thus would only
produce H in the very early phase of the mass transfer, unless it is
mixed efficiently with the rest of the star which certainly for white
dwarfs is not expected.

Apart maybe from XTE~J0929-314 all observed spectra seem to be
consistent with white dwarf donors, at odds with the conclusions of
\citet{nrj86} and indeed with the inferred mass-transfer rates,
\textit{if these represent the average mass-transfer rate over an
astronomical timescale}.  In particular for the C/O spectra, where
there is no hint of He, C/O white dwarf donors seem the only
option. This is important, because according to this formation channel
the systems first evolve to very short orbital periods (of a few min.)
and then start mass transfer and evolve to longer periods again. The
progenitors of the C/O donor systems are thus very strong
gravitational wave sources for LISA \citep[e.g.][]{nel03b}. The known
UCXBs might already be guaranteed sources for LISA and thus be used as
verification sources for the instrument
\citep[see][]{phi02,nel05}. However, for many of the systems the
orbital periods still have to be determined. This should therefore
have a high priority.

A second ingredient that will become very important in disentangling
the formation of UCXBs is the relative number of C/O vs He
systems. Although for He systems the formation channel is uncertain,
it is interesting to determine this ratio. For instance, it is telling
that the fact that currently the majority of the systems with known
chemical composition have C/O donors, while in the very related family
of AM CVn stars \citep[ultra-compact binaries which have white dwarf
instead of neutron star accretors, see][for a recent review]{nel05}
only He donors are known. Based on the stability of mass transfer at
the very short periods only low-mass donors are expected to be present
\citep[e.g.][]{npv+00,ynh02} which would preferentially select He
donors. \citet{bt04} simulated the Galactic population of UCXBs and
predict 60 per cent of the donors to be He rich. Apparently in UCXBs
there is an evolutionary or an observational bias towards C/O
donors. Only better statistics can tell. One of the important tools
here will be optical spectroscopy of UCXB transients when they are in
outburst.

Another important aspect of UCXBs is the distribution of persistent vs
transient systems (see Table~\ref{tab:properties}). Evolutionary
calculations combined with disc instability models predicts that for
UCXBs that have (semi) degenerate donors systems with periods longer
than about 30 min are transient, while shorter period systems are
persistent X-ray sources \citep[see][]{db03}. It is already clear that
the observed systems do not follow this prediction (2 or 3 persistent
systems have periods above 30 min), suggesting either incompleteness
in the disc instability model or a different formation channel, which
produces higher mass transfer rates at long periods. We note however,
that the classical division between transient and persistent sources
is beginning to be complicated by systems such as XB~1905+00 which
after 11 years of ``persistent'' emission has returned to quiescence. 

All UCXBs in the field for which we know the orbital period have
periods between 40 and 50 min, except perhaps 4U~1543-624. On
evolutionary grounds most systems are expected at long periods
\citep[e.g.][]{db03}, but the mass transfer rates are also lower,
making them fainter. In addition, the transient nature of longer
period systems might make them easier to find. Still, the period
distribution of UCXBs looks quite different from the distribution of
periods in AM CVn systems, where 11 out of the 13 systems with known
periods have periods below 40 min
\citep[see][]{nel05,wwr05,ahh+05}. However, the expected total active
population of UCXBs in the field is small \citep[e.g.][]{bt04}, so
small number statistics may dominate this discussion.

Finally, when it will be possible to obtain high quality spectra and
obtain detailed, quantitative chemical compositions of the accretion
disc through reliable spectral modelling we can use these to study the
interior composition of the donor stars at different phases in their
evolution which will open a new era in the study of stellar structure.

\section{Conclusions}\label{conclusions}

We have presented optical spectroscopy of (candidate) UCXBs and
conclude that with the current (lack of) understanding of the spectra
of hydrogen-poor accretion disc spectra we cannot make quantitative
statements about the detailed chemical abundances of the donors stars
in UCXBs, only determine the global chemical composition. We have
obtained the first optical spectrum of a He dominated disc in an UCXB,
XB~1916-05, implying a He-rich donor in this system. A simple LTE
model suggests that the N abundance in this systems is strongly
enhanced. We furthermore confirm the C/O nature of the donor stars in
4U~0614+09 and 4U~1626-67, and suggest the donor in XTE~J0929-314 is a
C/O white dwarf as well. Within the current sample the majority of the
donors are C/O rich. This suggests that in UCXBs there is an
evolutionary or observational bias towards C/O donors. Phase resolved
spectroscopy of 4U~0614+09 does not reveal a clear periodicity,
although there seem to be line variations at short time scales. We did
not detect the optical counterpart of XB~1905+00 and found that the
candidate UCXB 4U~1556-60 shows strong H emission and thus is not an
UCXB but an ordinary low-mass X-ray binary.

\section*{Acknowledgments}

Many thanks to Tom Marsh, for the use of his LTE emission line model
and MOLLY and to Gijs Roelofs for the use of his trail movie code. We
are thankful to Peter van Hoof and the National Institute of Standards
and Technology for compiling the atomic line lists we use.  GN and PGJ
are supported by the Netherlands Organisation of Scientific
Research. DS acknowledges a Smithsonian Astrophysical Observatory Clay
Fellowship. Based on observations obtained at the Gemini Observatory,
which is operated by the Association of Universities for Research in
Astronomy, Inc., under a cooperative agreement with the NSF on behalf
of the Gemini partnership: the National Science Foundation (United
States), the Particle Physics and Astronomy Research Council (United
Kingdom), the National Research Council (Canada), CONICYT (Chile), the
Australian Research Council (Australia), CNPq (Brazil) and CONICET
(Argentina)

\bibliography{journals,binaries} 

\begin{thebibliography}{}

\bibitem[\protect\citeauthoryear{{Anderson}, {Haggard}, {Homer}, {Joshi},
  {Margon}, {Silvestri}, {Szkody}, {Wolfe}, {Agol} et~al.,}{{Anderson}
  et~al.}{2005}]{ahh+05}
{Anderson} S.~F.,  {Haggard} D.,  {Homer} L.,  {Joshi} N.~R.,  {Margon} B.,
  {Silvestri} N.~M.,  {Szkody} P.,  {Wolfe} M.~A.,  {Agol} E.,    et~al., 2005,
  \aj, 130, 2230

\bibitem[\protect\citeauthoryear{{Bassa}, {Jonker}, {in 't Zand} \&
  {Verbunt}}{{Bassa} et~al.}{2006}]{bji+06}
{Bassa} C.~G.,  {Jonker} P.~G.,  {in 't Zand} J.~J.~M.,    {Verbunt} F.,  2006,
  \aap, p. in press

\bibitem[\protect\citeauthoryear{{Belczynski} \& {Taam}}{{Belczynski} \&
  {Taam}}{2004}]{bt04}
{Belczynski} K.,  {Taam} R.~E.,  2004, \apj, 603, 690

\bibitem[\protect\citeauthoryear{{Castro-Tirado}, {Caccianiga}, {Gorosabel},
  {Kilmartin}, {Tristram}, {Yock}, {Sanchez-Fernandez} \&
  {Alcoholado-Feltstrom}}{{Castro-Tirado} et~al.}{2002}]{ccg+02}
{Castro-Tirado} A.~J.,  {Caccianiga} A.,  {Gorosabel} J.,  {Kilmartin} P.,
  {Tristram} P.,  {Yock} P.,  {Sanchez-Fernandez} C.,    {Alcoholado-Feltstrom}
  M.~E.,  2002, \iaucirc, 7895, 1

\bibitem[\protect\citeauthoryear{{Chevalier} \& {Ilovaisky}}{{Chevalier} \&
  {Ilovaisky}}{1985}]{ci85}
{Chevalier} C.,  {Ilovaisky} S.~A.,  1985, Space Science Reviews, 40, 443

\bibitem[\protect\citeauthoryear{{Cowley}, {Hutchings} \& {Crampton}}{{Cowley}
  et~al.}{1988}]{chc88}
{Cowley} A.~P.,  {Hutchings} J.~B.,    {Crampton} D.,  1988, \apj, 333, 906

\bibitem[\protect\citeauthoryear{{Deloye} \& {Bildsten}}{{Deloye} \&
  {Bildsten}}{2003}]{db03}
{Deloye} C.~J.,  {Bildsten} L.,  2003, \apj, 598, 1217

\bibitem[\protect\citeauthoryear{{Farinelli}, {Frontera}, {Masetti}, {Amati},
  {Guidorzi}, {Orlandini}, {Palazzi}, {Parmar}, {Stella}, {Van der Klis} \&
  {Zhang}}{{Farinelli} et~al.}{2003}]{ffm+03}
{Farinelli} R.,  {Frontera} F.,  {Masetti} N.,  {Amati} L.,  {Guidorzi} C.,
  {Orlandini} M.,  {Palazzi} E.,  {Parmar} A.~N.,  {Stella} L.,  {Van der Klis}
  M.,    {Zhang} S.~N.,  2003, \aap, 402, 1021

\bibitem[\protect\citeauthoryear{{Galloway}, {Chakrabarty}, {Muno} \&
  {Savov}}{{Galloway} et~al.}{2001}]{gcm+01}
{Galloway} D.~K.,  {Chakrabarty} D.,  {Muno} M.~P.,    {Savov} P.,  2001,
  \apjl, 549, L85

\bibitem[\protect\citeauthoryear{{Galloway}, {Morgan}, {Remillard} \&
  {Chakrabarty}}{{Galloway} et~al.}{2002}]{gmr+02}
{Galloway} D.~K.,  {Morgan} E.~H.,  {Remillard} R.~A.,    {Chakrabarty} D.,
  2002, \iaucirc, 7900, 2

\bibitem[\protect\citeauthoryear{{Giles}, {Greenhill}, {Hill} \&
  {Sanders}}{{Giles} et~al.}{2005}]{ggh+05}
{Giles} A.~B.,  {Greenhill} J.~G.,  {Hill} K.~M.,    {Sanders} E.,  2005,
  \mnras, 361, 1180

\bibitem[\protect\citeauthoryear{{Homer}, {Anderson}, {Wachter} \&
  {Margon}}{{Homer} et~al.}{2002}]{haw+02}
{Homer} L.,  {Anderson} S.~F.,  {Wachter} S.,    {Margon} B.,  2002, \aj, 124,
  3348

\bibitem[\protect\citeauthoryear{{Horne}}{{Horne}}{1986}]{hor86}
{Horne} K.,  1986, \pasp, 98, 609

\bibitem[\protect\citeauthoryear{{in't Zand}, {Cumming}, {van der Sluys},
  {Verbunt} \& {Pols}}{{in't Zand} et~al.}{2005}]{icv05}
{in't Zand} J.~J.~M.,  {Cumming} A.,  {van der Sluys} M.~V.,  {Verbunt} F.,
  {Pols} O.~R.,  2005, \aap, 441, 675

\bibitem[\protect\citeauthoryear{Jonker, Bassa, Nelemans, Juett, Brown \&
  Chakrabarty}{Jonker et~al.}{2006}]{jbn+06}
Jonker P.,  Bassa C.,  Nelemans G.,  Juett A.,  Brown E.,    Chakrabarty D.,
  2006, \mnras, in press

\bibitem[\protect\citeauthoryear{Juett, Psaltis \& Chakrabarty}{Juett
  et~al.}{2001}]{jpc00}
Juett A.~M.,  Psaltis D.,    Chakrabarty D.,  2001, \apjl, 560, L59

\bibitem[\protect\citeauthoryear{{Markwardt} \& {Swank}}{{Markwardt} \&
  {Swank}}{2002}]{ms02a}
{Markwardt} C.~B.,  {Swank} J.~H.,  2002, \iaucirc, 7867, 1

\bibitem[\protect\citeauthoryear{Marsh, Horne \& Rosen}{Marsh
  et~al.}{1991}]{mhr91}
Marsh T.~R.,  Horne K.,    Rosen S.,  1991, ApJ, 366, 535

\bibitem[\protect\citeauthoryear{{Middleditch}, {Mason}, {Nelson} \&
  {White}}{{Middleditch} et~al.}{1981}]{mmn81}
{Middleditch} J.,  {Mason} K.~O.,  {Nelson} J.~E.,    {White} N.~E.,  1981,
  \apj, 244, 1001

\bibitem[\protect\citeauthoryear{{Morales-Rueda}, {Marsh}, {Steeghs},
  {Unda-Sanzana}, {Wood} \& {North}}{{Morales-Rueda} et~al.}{2003}]{mms+03}
{Morales-Rueda} L.,  {Marsh} T.~R.,  {Steeghs} D.,  {Unda-Sanzana} E.,  {Wood}
  J.~H.,    {North} R.~C.,  2003, \aap, 405, 249

\bibitem[\protect\citeauthoryear{{Motch}, {Pakull}, {Mouchet} \&
  {Beuermann}}{{Motch} et~al.}{1989}]{mpm+89}
{Motch} C.,  {Pakull} M.~W.,  {Mouchet} M.,    {Beuermann} K.,  1989, \aap,
  219, 158

\bibitem[\protect\citeauthoryear{{Nelemans}}{{Nelemans}}{2003}]{nel03b}
{Nelemans} G.,  2003, in Centrella J.,  ed., The Astrophysics of Gravitational
  Wave Sources Vol.~686 of AIP conf. proc., {Galactic binaries as sources of
  Gravitational waves}.
AIP, New York, p.~263

\bibitem[\protect\citeauthoryear{{Nelemans}}{{Nelemans}}{2005}]{nel05}
{Nelemans} G.,  2005, in ASP Conf. Ser. 330: The Astrophysics of Cataclysmic
  Variables and Related Objects {AM CVn stars}.
p.~27

\bibitem[\protect\citeauthoryear{Nelemans \& Jonker}{Nelemans \&
  Jonker}{2005}]{nj04}
Nelemans G.,  Jonker P.,  2005, in Antonelli L.,  Burderi L.,  D'Antona F.,
  Di~Salvo T.,  Israel G.,  Piersanti L.,  Straniero O.,   Tornamb\`e A.,  eds,
  Interacting binaries AIP Conf. Proc., Optical spectroscopy of (candidate)
  ultra-compact x-ray binaries.
AIP, New York, p.~396

\bibitem[\protect\citeauthoryear{Nelemans \& Jonker}{Nelemans \&
  Jonker}{2006}]{nj06}
Nelemans G.,  Jonker P.,  2006, New Ast. Rev., submitted

\bibitem[\protect\citeauthoryear{Nelemans, Jonker, Marsh \& van~der
  Klis}{Nelemans et~al.}{2004}]{njm+04}
Nelemans G.,  Jonker P.~G.,  Marsh T.~R.,    van~der Klis M.,  2004, \mnras,
  348, L7

\bibitem[\protect\citeauthoryear{Nelemans, Portegies~Zwart, Verbunt \&
  Yungelson}{Nelemans et~al.}{2001}]{npv+00}
Nelemans G.,  Portegies~Zwart S.~F.,  Verbunt F.,    Yungelson L.~R.,  2001,
  A\&A, 368, 939

\bibitem[\protect\citeauthoryear{{Nelson}, {Rappaport} \& {Joss}}{{Nelson}
  et~al.}{1986}]{nrj86}
{Nelson} L.~A.,  {Rappaport} S.~A.,    {Joss} P.~C.,  1986, \apj, 304, 231

\bibitem[\protect\citeauthoryear{{O'Brien}, {Jonker}, {Dhillon}, {Nelemans},
  {Stil}, {van der Klis} \& {Marsh}}{{O'Brien} et~al.}{2005}]{ojd+05}
{O'Brien} K.,  {Jonker} P.~G.,  {Dhillon} V.,  {Nelemans} G.,  {Stil} M.,  {van
  der Klis} M.,    {Marsh} T.~R.,  2005, \aap, in prep

\bibitem[\protect\citeauthoryear{Phinney}{Phinney}{2002}]{phi02}
Phinney E.,  2002, Technical report, LISA Science Requirements.
Caltech

\bibitem[\protect\citeauthoryear{{Podsiadlowski}, {Rappaport} \&
  {Pfahl}}{{Podsiadlowski} et~al.}{2002}]{prp02}
{Podsiadlowski} P.,  {Rappaport} S.,    {Pfahl} E.~D.,  2002, \apj, 565, 1107

\bibitem[\protect\citeauthoryear{{Remillard}, {Swank} \&
  {Strohmayer}}{{Remillard} et~al.}{2002}]{rss02}
{Remillard} R.~A.,  {Swank} J.,    {Strohmayer} T.,  2002, \iaucirc, 7893, 1

\bibitem[\protect\citeauthoryear{Schulz, Chakrabarty, Marshall, Canizares, Lee
  \& J.}{Schulz et~al.}{2001}]{sch+01}
Schulz N.~S.,  Chakrabarty D.,  Marshall H.~L.,  Canizares C.~R.,  Lee L.~C.,
   J. H.,  2001, \apj, 563, 941

\bibitem[\protect\citeauthoryear{van~den Heuvel}{van~den Heuvel}{1983}]{heu83}
van~den Heuvel E. P.~J.,  1983, in Lewin W. H.~G.,  van~den Heuvel E. P.~J.
  eds, , Accretion-driven stellar X-ray sources.
CUP, Cambridge, pp 303--341

\bibitem[\protect\citeauthoryear{{van der Sluys}, {Verbunt} \& {Pols}}{{van der
  Sluys} et~al.}{2005}]{svp05a}
{van der Sluys} M.~V.,  {Verbunt} F.,    {Pols} O.~R.,  2005, \aap, 440, 973

\bibitem[\protect\citeauthoryear{{van Paradijs} \& {McClintock}}{{van Paradijs}
  \& {McClintock}}{1994}]{vm94}
{van Paradijs} J.,  {McClintock} J.~E.,  1994, \aap, 290, 133

\bibitem[\protect\citeauthoryear{{Verbunt}}{{Verbunt}}{2005}]{ver04}
{Verbunt} F.,  2005, in Antonelli L.,  Burderi L.,  D'Antona F.,  Di~Salvo T.,
  Israel G.,  Piersanti L.,  Straniero O.,   Tornamb\`e A.,  eds, Interacting
  binaries AIP Conf. Proc., {X-ray sources in globular clusters}.
AIP, New York, p.~30

\bibitem[\protect\citeauthoryear{Verbunt \& van~den Heuvel}{Verbunt \& van~den
  Heuvel}{1995}]{vh95}
Verbunt F.,  van~den Heuvel E.~P.~J.,  1995, in Lewin W. H.~G.,  van
  Paradijs~J.  van~den Heuvel E. P.~J. eds, , X-ray Binaries.
Cambridge: Cambridge Univ.~Press, pp 457--494

\bibitem[\protect\citeauthoryear{{Walter}, {Mason}, {Clarke}, {Halpern},
  {Grindlay}, {Bowyer} \& {Henry}}{{Walter} et~al.}{1982}]{wmc+82}
{Walter} F.~M.,  {Mason} K.~O.,  {Clarke} J.~T.,  {Halpern} J.,  {Grindlay}
  J.~E.,  {Bowyer} S.,    {Henry} J.~P.,  1982, \apjl, 253, L67

\bibitem[\protect\citeauthoryear{{Warner} \& {Woudt}}{{Warner} \&
  {Woudt}}{2002}]{ww02}
{Warner} B.,  {Woudt} P.~A.,  2002, \pasp, 114, 129

\bibitem[\protect\citeauthoryear{{Werner}, {Nagel}, {Dreizler} \&
  {Rauch}}{{Werner} et~al.}{2004}]{wnd+04}
{Werner} K.,  {Nagel} T.,  {Dreizler} S.,    {Rauch} T.,  2004, in Tovmassian
  G.,  Sion E.,  eds, Compact binaries in The Galaxy and beyond Vol.~20 of
  RevMexAA (SC), {Modeling of Oxygen-Neon Dominated Accretion Disks in
  Ultracompact X-ray Binaries: 4U 1626-67}.
pp 146--147

\bibitem[\protect\citeauthoryear{{Werner}, {Nagel}, {Rauch}, Hammer \&
  {Dreizler}}{{Werner} et~al.}{2006}]{wnr+06}
{Werner} K.,  {Nagel} T.,  {Rauch} T.,  Hammer N.,    {Dreizler} S.,  2006,
  \aap, in press

\bibitem[\protect\citeauthoryear{{White} \& {Swank}}{{White} \&
  {Swank}}{1982}]{ws82}
{White} N.~E.,  {Swank} J.~H.,  1982, \apjl, 253, L61

\bibitem[\protect\citeauthoryear{{Woudt}, {Warner} \& {Rykoff}}{{Woudt}
  et~al.}{2005}]{wwr05}
{Woudt} P.~A.,  {Warner} B.,    {Rykoff} E.,  2005, \iaucirc, 8531, 3

\bibitem[\protect\citeauthoryear{{Yungelson}, {Nelemans} \& {van den
  Heuvel}}{{Yungelson} et~al.}{2002}]{ynh02}
{Yungelson} L.~R.,  {Nelemans} G.,    {van den Heuvel} E.~P.~J.,  2002, \aap,
  388, 546

\end{thebibliography}
\bibliographystyle{mn2e}

\label{lastpage}

\end{document}